\documentclass[reprint,showpacs,showkeys,amsmath,amssymb,superscriptaddress,aps,pra,longbibliography]{revtex4-1}
\usepackage{graphicx}% Include figure files
\usepackage{float}

\usepackage{lipsum}
\usepackage{epstopdf}
\usepackage{bm}
\usepackage{amssymb}
\usepackage{amsmath}
\usepackage{bbold}
\usepackage[colorlinks=true, linkcolor=blue]{hyperref}
\usepackage{ulem}

\usepackage{xcolor}
\usepackage{soul}
\usepackage[utf8]{inputenc}
\DeclareUnicodeCharacter{2212}{-}
\DeclareUnicodeCharacter{B0}{\textdegree}
\usepackage{chemformula}
\DeclareUnicodeCharacter{0301}{\'{e}}
\usepackage{gensymb}
\usepackage{graphicx}
\usepackage{textcomp, gensymb}
\usepackage{tikz}
\usepackage{float}
\usepackage{silence}
\DeclareUnicodeCharacter{0301}{\'{e}}
\DeclareUnicodeCharacter{2218}{\'{e}}

\begin{document}

\title{Enhancement of spin Hall angle by an order of magnitude via Cu intercalation in MoS$_2$/CoFeB heterostructures}

\author {Abhisek Mishra} 
\affiliation {Laboratory for Nanomagnetism and Magnetic Materials, School of Physical Sciences, National Institute of Science Education and Research (NISER), An OCC of Homi Bhabha National Institute (HBNI), Jatni 752050, India.}

\author {Pritam Das}
\affiliation {School of Advanced Material Engineering, Kookmin University, Seoul 02707, Republic of Korea}

\author{Rupalipriyadarsini Chhatoi}
\affiliation {Laboratory for Nanomagnetism and Magnetic Materials, School of Physical Sciences, National Institute of Science Education and Research (NISER), An OCC of Homi Bhabha National Institute (HBNI), Jatni 752050, India.}

\author{Soubhagya Dash}
\affiliation {Laboratory for Nanomagnetism and Magnetic Materials, School of Physical Sciences, National Institute of Science Education and Research (NISER), An OCC of Homi Bhabha National Institute (HBNI), Jatni 752050, India.}

\author{Shubhransu Sahoo}
\affiliation {Laboratory for Nanomagnetism and Magnetic Materials, School of Physical Sciences, National Institute of Science Education and Research (NISER), An OCC of Homi Bhabha National Institute (HBNI), Jatni 752050, India.}

\author{Kshitij Singh Rathore}
\affiliation {Laboratory for Nanomagnetism and Magnetic Materials, School of Physical Sciences, National Institute of Science Education and Research (NISER), An OCC of Homi Bhabha National Institute (HBNI), Jatni 752050, India.}

\author{Pil-Ryung Cha}
\affiliation {School of Advanced Material Engineering, Kookmin University, Seoul 02707, Republic of Korea}

\author{Seung-Cheol Lee}
\affiliation{Korea Institute of Science and Technology, Seoul, Republic of Korea}

\author{Satadeep Bhattacharjee}
\email{s.bhattacharjee@ikst.res.in}
\affiliation {Indo-Korea Science and Technology Center (IKST), Bangalore, India}

\author{Subhankar Bedanta}
\email{sbedanta@niser.ac.in}
\affiliation {Laboratory for Nanomagnetism and Magnetic Materials, School of Physical Sciences, National Institute of Science Education and Research (NISER), An OCC of Homi Bhabha National Institute (HBNI), Jatni 752050, India.}

\affiliation{Center for Interdisciplinary Sciences (CIS), National Institute of Science Education and Research (NISER), An OCC of Homi Bhabha National Institute (HBNI), Jatni 752050, India.}

\date{\today}

\begin{abstract}
Transition metal dichalcogenides (TMDs) are a novel class of quantum materials with significant potential in spintronics, optoelectronics, valleytronics, and opto-valleytronics. TMDs exhibit strong spin-orbit coupling, enabling efficient spin-charge interconversion, which makes them ideal candidates for spin-orbit torque-driven spintronic devices. In this study, we investigated the spin-to-charge conversion through ferromagnetic resonance in MoS$_2$/Cu/CoFeB heterostructures with varying Cu spacer thicknesses. The conversion efficiency, quantified by the spin Hall angle, was enhanced by an order of magnitude due to Cu intercalation. Magneto-optic Kerr effect microscopy confirmed that Cu did not significantly modify the magnetic domains, indicating its effectiveness in decouplingMoS$_2$ from CoFeB. This decoupling preserves the spin-orbit coupling (SOC) of MoS$_2$ by mitigating the exchange interaction with CoFeB, as proximity to localized magnetization can alter the electronic structure and SOC. First-principles calculations revealed that Cu intercalation notably enhances the spin Berry curvature and spin Hall conductivity, contributing to the increased spin Hall angle. This study demonstrates that interface engineering of ferromagnet/TMD-based heterostructures can achieve higher spin-to-charge conversion efficiencies, paving the way for advancements in spintronic applications. 

\end{abstract}



\maketitle
\section{Introduction}
Achieving high-density spin currents in materials with high spin-orbit coupling (SOC) is critical for advancing spintronic applications \cite{hirohata2020review,soumyanarayanan2016emergent,ryu2020current}. Among the various mechanisms explored, spin pumping has emerged as a highly effective approach \cite{tserkovnyak2002spin}. This technique generates substantial spin current densities in high-SOC materials adjacent to a ferromagnet (FM) or ferrimagnet (FiM) with precessing magnetization. Spin pumping is particularly advantageous in heterostructures comprising metallic, insulating, or semiconducting high-SOC materials, addressing impedance mismatch and enhancing device feasibility. Traditionally, the spin Hall effect (SHE) and its Onsager reciprocal, the inverse spin Hall effect (ISHE), have been the primary mechanisms for bidirectional spin-charge conversion, driven by electron scattering in three-dimensional high-SOC materials such as heavy metals (e.g., Pt, Pd, W, and Ta), antiferromagnetic materials (e.g., IrMn, Mn$_2$Au, Mn$_3$Ga, Mn$_3$Sn), and topological insulators (e.g., Bi$_2$Se$_3$, Bi$_2$Te$_3$) \cite{hirsch1999spin,saitoh2006conversion,roy2021spin,singh2020inverse,singh2020large,kimata2019magnetic,singh2019inverse,jamali2015giant}. Recently, transition metal dichalcogenides (TMDs) have been identified as promising quantum materials for enhanced spin-to-charge conversion efficiency due to their high SOC \cite{sierra2021van}. Among TMDs, molybdenum disulfide (MoS$_2$) stands out for its stability, robustness, and unique electronic properties, which vary with thickness and mechanical strain \cite{liu2020spintronics,kumar2022next,choi2017recent,akinwande2014two}. In MoS$_2$/FM heterostructures, interface effects significantly enhance global anisotropy, and MoS$_2$ has demonstrated remarkable spin-to-charge conversion voltages and efficiencies \cite{xie2019giant,thiruvengadam2022anisotropy,mishra2024spin,mendes2018efficient,husain2018spin,bansal2019extrinsic}. Furthermore, several mechanisms have been suggested to improve the efficiency of spin-to-charge conversion in FM/high-SOC systems \cite{du2014enhancement,longo2021spin,obstbaum2016tuning,chen2024tuning,choi2024non}. Among them, the utilization of interlayers has been an effective approach. These interlayers act as barriers, preserving the SOC of high-SOC materials by mitigating proximity-induced magnetic effects and tuning spin-dependent interfacial resistivity and spin injection. In this context, Cu emerges as a promising interlayer material due to its long spin diffusion length ($\sim$ 350 nm), low SOC, and ability to control spin transmissivity in spintronic devices \cite{yakata2006temperature}.  

Here, we report a significant enhancement—by an order of magnitude—in the spin Hall angle (SHA) of MoS$_2$ through Cu intercalation in MoS$_2$/CoFeB heterostructures.  Magneto-optic Kerr microscopy reveals that Cu effectively decouples MoS$_2$ from CoFeB, thereby preserving the SOC of MoS$_2$. These experimental results are supported by density functional theory (DFT) calculations, which highlight the role of Cu in enhancing spin-to-charge conversion efficiency.
\section{Experimental details}
Thin films of MoS$_2$(5 nm)/Cu ($t_{Cu}$ nm)/CoFeB (9 nm)/Al$_2$O$_3$(3 nm) with varying Cu thicknesses were fabricated using a high-vacuum multi-deposition chamber (Mantis Deposition Ltd., UK) with a base pressure below 4 $\times$10$^{-8}$ mbar. MoS$_2$, Co$_{40}$Fe$_{40}$B$_{20}$ (CoFeB) and Al$_2$O$_3$ films were deposited from commercially available stoichiometric targets. Cu and CoFeB were deposited via \textit{dc} magnetron sputtering whereas MoS$_2$ and Al$_2$O$_3$ were deposited by \textit{rf} sputtering. The samples were grown on Si (100) substrates with a 300 nm thick SiO$_2$ layer. Samples, labeled M1 to M7, correspond to Cu thicknesses ($t_{Cu}$) of 0, 0.65, 2, 3, 5, 7, and 10 nm, respectively. The reference sample, CoFeB (9 nm)/Al$_2$O$_3$ (3 nm), was designated as M0. Magnetization dynamics and inverse spin Hall effect (ISHE) measurements were conducted using ferromagnetic resonance (FMR). Saturation magnetization was measured with a SQUID-based magnetometer (MPMS 3, Quantum Design). Room-temperature longitudinal magneto-optic Kerr effect (MOKE) microscopy and magnetometry were employed to study magnetic domains and hysteresis loops, respectively (see Fig. S7 of Supplemental Material \cite{SM_PRA1})).
\nocite{kittel1948theory}
\nocite{brataas2002spin}
\nocite{mosendz2010detection}
\nocite{kresse1996efficiency}
\nocite{perdew1996generalized}
\nocite{daalderop1990first}
\nocite{liechtenstein1987local}
Further details of the experimental protocols can be found in the Supplemental Material \cite{SM_PRA1}.

\section{Results and discussion}
\begin{figure}[ht]
	\centering
	\includegraphics[width=0.5\textwidth]{"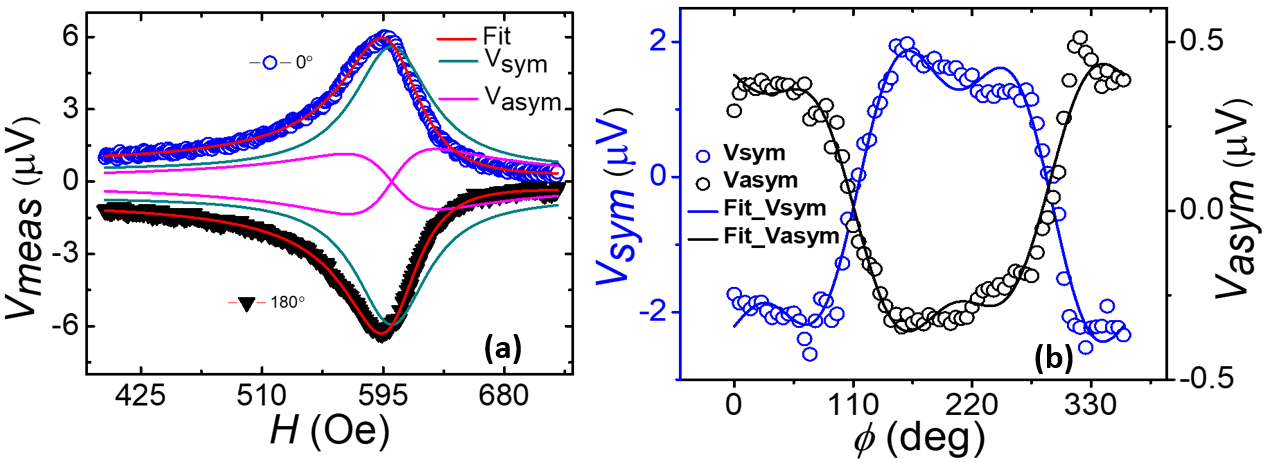"}
	\caption{(a) Measured \textit{dc} voltage signals for 0$^\circ$ (open blue symbols) and 180$^\circ$ (solid black triangles) for sample M5 are shown as open circles. The solid red line is the fit to equation (1). The green and magenta lines are the $V_{sym}$ and $V_{asym}$ components of the voltage. (b) $\phi$ dependent $V_{sym}$ and $V_{asym}$ for samples M5, which are fitted to eqns. (2) and (3), respectively}

	\label{fig:figure-1}
\end{figure}
The presence of MoS$_2$ in the samples was confirmed through laser Raman spectroscopy, while the thickness and surface roughness were estimated using X-ray reflectivity (XRR). Detailed structural information obtained from Raman spectroscopy and XRR is provided in the Supplemental Material \cite{SM_PRA1}. The Gilbert damping parameter ($\alpha$) was extracted by fitting the resonance field ($H_{res}$) and linewidth ($\Delta H$) data obtained from ferromagnetic resonance (FMR) spectroscopy, as described in the Supplemental Material \cite{SM_PRA1}. The $\alpha$ values ($\times$10$^{-3}$) for samples M1–M7 are listed in Table I and are higher than that of the reference sample M0 (8.2$\pm$0.1), indicating potential spin pumping effects. However, other contributing mechanisms cannot be ruled out. To verify spin pumping, ISHE measurements were performed. In-plane angle-dependent ISHE measurements are particularly advantageous because they provide a direct, controlled, and quantitative probe of spin-charge conversion with high sensitivity. Figure 1(a) shows the variation in measured ISHE voltage ($V_{dc}$) as a function of applied field (\textit{H}) for sample M5 at in-plane angles ($\phi$) of 0$^\circ$ and 180$^\circ$. The symmetric (V$_{sym}$) and antisymmetric (V$_{asym}$) components of $V_{dc}$ were separated by fitting the data to the following Lorentzian equation, \cite{conca2017lack},
\begin{equation}
\begin{aligned}
    V_{dc} = V_{sym} \frac{(\Delta H)^2}{(H-H_{res})^2+(\Delta H)^2}+\\V_{asym} \frac{2 \Delta H (H - H_{res})}{(H-H_{res})^2+(\Delta H)^2}
    \end{aligned}
\end{equation}

\begin{figure}[ht]
	\centering
	\includegraphics[width=0.5\textwidth]{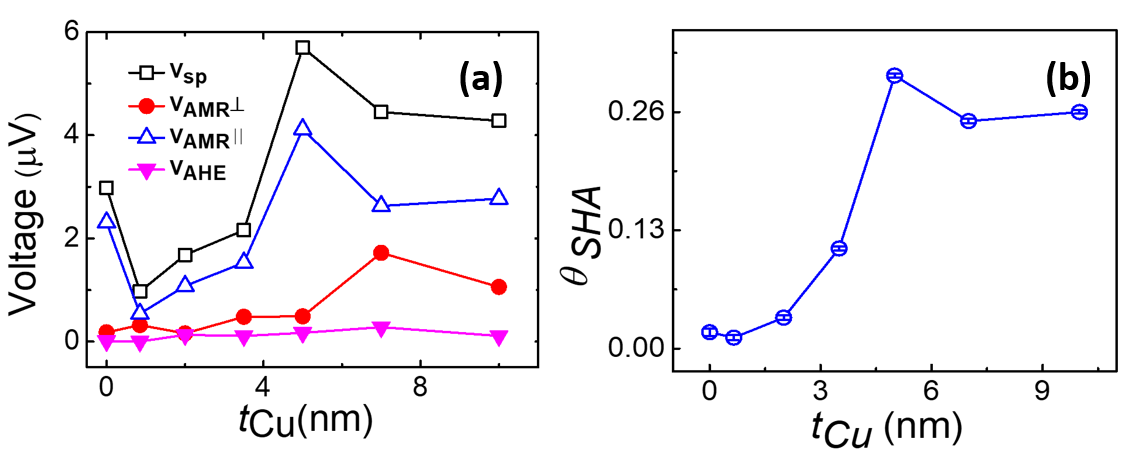}
	\caption{(a) The voltage contributions due to spin pumping and other spin rectification effects as a function of $t_{Cu}$. $V_{sp}$ shows a dominating contribution over other rectification effects in all the samples. (b) Variation of $\theta_{SHA}$ as a function of $t_{Cu}$.}
    \label{fig:figure-2}
\end{figure}
 Here, $\phi$ represents the angle between the direction of the measured voltage and the direction perpendicular to the applied \textit{H}. The sign reversal of $V_{dc}$ as $\phi$ changes from 0$^\circ$ to 180$^\circ$ clearly indicates the presence of spin pumping in the samples. In ISHE experiments, the detected voltage signal may also include contributions from parasitic spin rectification effects (SREs), which result from the nonlinear coupling between the dynamic resistance \textit{R}(t) influenced by the time-varying magnetic field \textit{H}(t) and dynamic current \textit{I}(t). SREs are primarily driven by mechanisms like the anomalous Hall effect (AHE) and anisotropic magnetoresistance (AMR). To quantify these contributions, angle-dependent ISHE measurements were conducted. Figures 1(b) and 1(c) show the angle-dependent $V_{sym}$ and $V_{asym}$ components for sample M5, respectively. The plots were fitted using the following equations \cite{conca2017lack}:

\begin{equation}
\begin{aligned}
{V_{sym}= V_{sp}cos^3(\phi + \phi_0)+V_{AHE}\ cos(\theta)}cos(\phi + \phi_0)\\
    + V_{sym}^{AMR \perp} cos 2(\phi + \phi_0)cos(\phi+ \phi_0)\\
    + V_{sym}^{AMR ||}sin2(\phi + \phi_0)cos(\phi+\phi_0)
    \end{aligned}
\end{equation}
\begin{equation}
\begin{aligned}
 V_{asym}=V_{AHE}\ sin (\theta) cos(\phi + \phi_0) + 
    \\V_{asym}^{AMR \perp} cos 2(\phi + \phi_0)cos(\phi + \phi_0)+ 
   \\ V_{asym}^{AMR ||}sin2(\phi + \phi_0)cos(\phi+\phi_0)
   \end{aligned}
 \end{equation}
An additional factor $\phi_0$ has been included to account for potential misalignment in sample positioning when defining the $\phi$ values during measurements. Here, $\theta$, the angle between the electric and magnetic fields of the applied microwave, is 90$^\circ$. The various voltage components derived from angle-dependent ISHE measurements are summarized in Table I.

\begin{table*}[t]
\caption{$\alpha$, $g_{eff}^{\uparrow\downarrow}$, \textit{R} and fitted parameters from the in-plane angle dependent ISHE measurements}

\centering
\begin{tabular}{ccccccccc} \label{table2}

Sample & $t_{Cu}$ (nm) & $V_{sp}$(V)$\times$$10^{-6}$ & $V_{AHE}$(V)$\times$$10^{-6}$ & $V_{AMR}^{\perp}$(V)$\times$$10^{-6}$ & $V_{AMR}^{||}$(V)$\times$$10^{-6}$    & $\alpha$($\times$10$^{-3}$)     & $g_{eff}^{\uparrow\downarrow}$(nm$^{-2}$)  &
\textit{R} ($\Omega$)
\\ \hline
M1     & 0 & 2.98 $\pm$ 0.08              & 0.18 $\pm$ 0.06              & 2.31 $\pm$ 0.08                       & 0.012 $\pm$ 0.001        & 11.40 $\pm$ 0.10       & 14.60 $\pm$ 0.02   & 362$\pm$0.72             \\ \hline
M2     & 0.65 & 0.97 $\pm$ 0.02           & 0.32 $\pm$ 0.01              & 0.54 $\pm$ 0.03                    & 0.002 $\pm$ 0.001           & 13.60 $\pm$ 0.10        & 24.70 $\pm$ 0.04      & 240$\pm$0.68         \\ \hline
M3     & 2 & 1.68 $\pm$ 0.05              & 0.16 $\pm$ 0.02              & 1.08 $\pm$ 0.06                       & 0.13 $\pm$ 0.03          & 13.20 $\pm$ 0.20         & 27.10 $\pm$ 0.03         & 75$\pm$0.67       \\ \hline
M4     & 3 & 2.16 $\pm$ 0.11              & 0.48 $\pm$ 0.04              & 1.53 $\pm$ 0.12                       & 0.11 $\pm$ 0.05           & 13.30 $\pm$ 0.10         & 28.30 $\pm$ 0.20     & 37$\pm$0.69          \\ \hline
M5     & 5 & 5.70 $\pm$ 0.12              & 0.49 $\pm$ 0.01              & 3.36 $\pm$ 0.13                       & 0.17 $\pm$ 0.06            & 13.01 $\pm$ 0.10         & 26.20 $\pm$ 0.20     & 30$\pm$0.66    \\ \hline
M6     & 7 & 4.45 $\pm$ 0.13              & 1.72 $\pm$ 0.05              & 2.63 $\pm$ 0.13                       & 0.28 $\pm$ 0.06           & 12.80 $\pm$ 0.10           & 25.50 $\pm$ 0.30      & 28$\pm$0.72    \\ \hline
M7     & 10 &  4.28 $\pm$ 0.05              & 1.06 $\pm$ 0.04              & 2.77 $\pm$ 0.05                       & 0.11 $\pm$ 0.02         & 12.90 $\pm$ 0.20        & 25.27 $\pm$ 0.20   & 26 $\pm$0.64      \\ 
\end{tabular}

\end{table*}

It has been observed that $V_{sp}$ dominates over the rectification effects, indicating a strong spin pumping phenomenon as shown in Fig. 2(a). The effective spin mixing conductance $g_{eff}^{\uparrow\downarrow}$, which governs the spin current flow across the interface, was calculated using the following expression \cite{zhu2019effective}:
\begin{equation}
 g_{eff}^{\uparrow\downarrow}=\frac{\Delta\alpha 4\pi M_{s}t_{CoFeB}}{g\mu_{B}} 
\end{equation}

 Here, \textit{$\Delta\alpha$} represents the change in damping from the reference CoFeB layer, \textit{$M_S$} is the saturation magnetization, \textit{$t_{CoFeB}$} is the thickness of the CoFeB layer, \textit{$\mu_{B}$}  is the Bohr magneton, and \textit{g} is the Landé g-factor. Saturation magnetization values ($M_S$) for samples M0–M7 were determined using SQUID magnetometry and found to be 801 $\pm$ 31, 840 $\pm$ 25, 790 $\pm$ 36, 833 $\pm$ 20, 843 $\pm$ 32, 820 $\pm$ 34, 848 $\pm$ 23 and 818 $\pm$ 28 emu/cc, respectively. The calculated $g_{eff}^{\uparrow\downarrow}$ values for samples M1–M7 are summarized in Table I. The $g_{eff}^{\uparrow\downarrow}$ values for the trilayer structures incorporating Cu were higher than those for the bilayer structures without Cu, suggesting that the Cu layer facilitates efficient spin transfer across the interface.

 The spin-to-charge conversion efficiency ($\theta_{SHA}$), also known as the spin Hall angle, was evaluated using the following expression \cite{ando2011inverse}:
 \begin{equation}
    \frac{V_{SP}}{R} = w\times \theta_{SHA} \lambda_{SD}tanh(\frac{t_{MoS_2}}{2 \lambda_{SD}}) {J}_s
\end{equation}
where \textit{$J_S$} is given by,
\begin{equation}
\begin{aligned}
    {J}_s \approx\frac{{g_{eff}^{\uparrow\downarrow}}{\gamma}^2 {h_{rf}}^2\hbar[4\pi M_s\gamma+\sqrt{(4\pi M_s\gamma)^2+16(\pi f)^2}]}{8\pi{\alpha}^2[{(4\pi M_s\gamma)^2+16(\pi f)^2}]}
    \end{aligned}
\end{equation}
Here, \textit{w} represents the coplanar waveguide (CPW) transmission linewidth, \textit{R} is the sample resistance measured via the four-probe method, and $\lambda_{SD}$ is the spin diffusion length. In our setup, the \textit{rf} field ({$\mu_0$}$h_{rf}$) is 0.5 Oe, and the transmission line width, \textit{w} is 200 $\mu$m. We have considered the spin diffusion length of 7.83 nm from our previous report \cite{mishra2024spin}. Fig. 2(b) shows the variation of $\theta_{SHA}$ as a function of $t_{Cu}$. The maximum $\theta_{SHA}$ value of 0.30$\pm$0.01 was observed at $\theta_{SHA}$=5 nm, beyond which it decreased. This value is an order of magnitude higher than that of the MoS$_2$/CoFeB bilayer (sample M1), which exhibited $\theta_{SHA}$=0.020$\pm$0.003.

\begin{figure*}[ht]
	\centering
	\includegraphics[width=1\textwidth]{"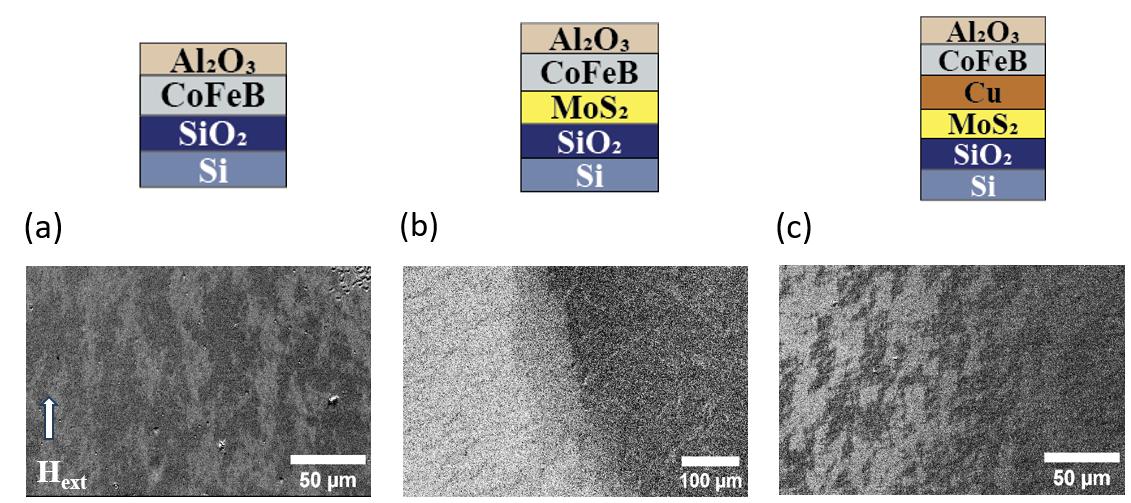"}
	\caption{Domain images captured for samples M0 (a), M1 (b) and M5 (c)  near to coercivity. Magnetic field direction shown in (a) is valid for all the domain images.}
	\label{fig:figure-3}
\end{figure*}
 The resistance of the heterostructures decreases gradually due to Cu intercalation and varies from 362$\pm$0.45 to 26$\pm$0.09 $\Omega$ from sample M1 to M7. With increasing Cu thickness, the Cu layer starts to act as a good spin transmitter with minimal dissipation and reduced spin flipping, preserving the spin polarization. This leads to an increase in the spin Hall angle. At $t_{Cu}$ $\geq$ 5 nm, the Cu layer becomes sufficiently thick to support spin accumulation. Spin accumulation refers to the buildup of spin-polarized electrons at the interfaces, enabled by a thickness that allows spins to diffuse without substantial spin-flip scattering. This suggests that the spin Hall angle of the Cu interlayer generates a positive voltage response, reducing the net spin Hall angle \cite{wang2014scaling}. In the MoS$_2$/CoFeB bilayer structure, the direct interface likely causes significant spin dissipation and the formation of mixed electronic states, both of which impede efficient spin accumulation. By introducing a Cu interlayer, these effects are mitigated, creating a cleaner pathway for spin currents to reach the MoS$_2$ layer and improving overall spin transport efficiency.  
 
\begin{figure*}[htp]
	\centering
	\includegraphics[width=0.8\textwidth]{"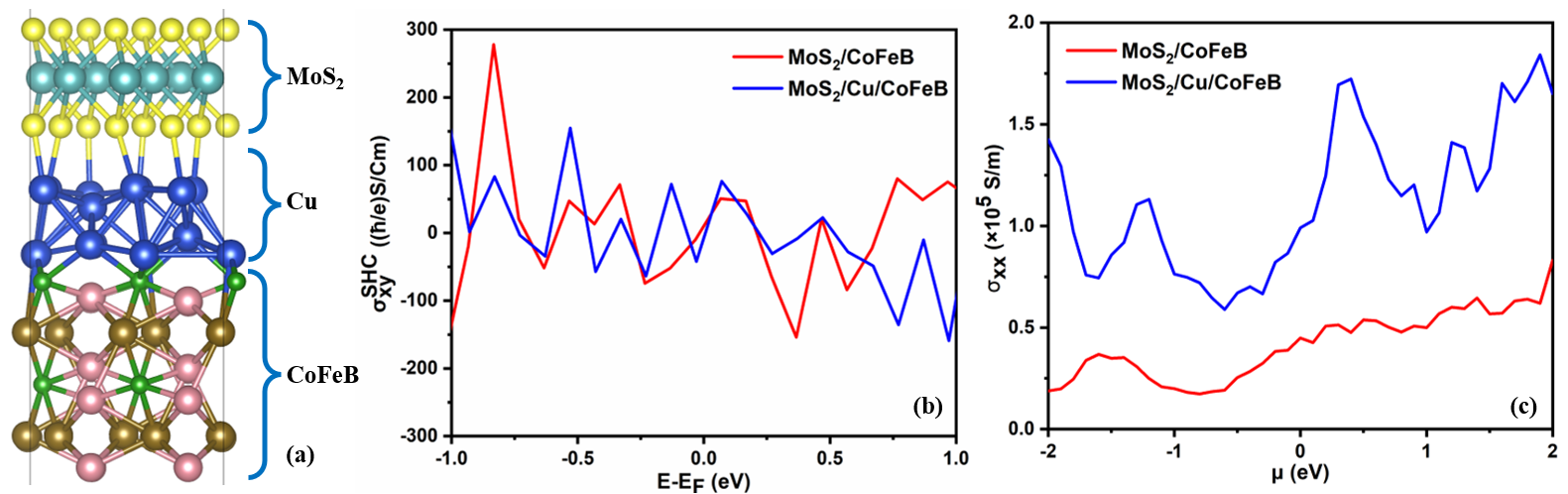"}
	\caption{ (a) Structural model for the MoS$_2$/Cu/CoFeB considered for the DFT-Wannier calculations for the spin transport. (b) Spin Berry conductivity as a function of energy relative to the Fermi energy, (c) the longitudinal charge
conductivity as a function of the chemical potential.}
	\label{fig:figure-4}
\end{figure*}
To further validate the interfacial decoupling in the MoS$_2$/Cu/CoFeB trilayer, we performed MOKE microscopy. When MoS$_2$ is directly interfaced with CoFeB, we observe larger magnetic domains, suggesting that magnetization reversal is dominated by domain wall nucleation and motion \cite{chowdhury2014controlling}. This is consistent with interfacial hybridization effects reported in earlier findings \cite{thiruvengadam2022anisotropy}. This can be observed from the domain image of M1 (MoS$_2$/CoFeB) at 0$^\circ$, shown in Fig. 3(b).  However, when Cu is intercalated between MoS$_2$ and CoFeB, it displays fuzzy domains shown in Fig. 3 (c). This is similar to that in the reference bare CoFeB layer (Fig. 3 (a)). 

It is to be noted that all our samples, with or without a Cu insertion layer, exhibit in-plane magnetic anisotropy. The effective anisotropy energy density $K_{eff}$ has been quantified via SQUID magnetometry (see Supplemental Material \cite{SM_PRA1}). The $K_{eff}$ for samples M0, M1 and M5 are (6.97$\pm$0.28)$\times$10$^6$, (5.67$\pm$0.21)$\times$10$^6$ and (6.65$\pm$0.32)$\times$10$^6$erg/cc, respectively. Therefore, the effective anisotropy is found to be enhanced with Cu intercalation. Similar trend has also been found from the DFT calculations (to be discussed later). 

\section{Theoretical calculations}
To understand the role of the Cu spacer layer in the spin-pumping we performed first-principles based calculation to estimate the spin Hall angles (SHA) in MoS$_2$/CoFeB and MoS$_2$/Cu/CoFeB systems within the framework of Density Functional Theory (DFT). The detailed methodology of these calculations is mentioned in the Supplemental Material \cite{SM_PRA1}. 
The spin Hall angle is given by \cite{qu2014self},

\begin{equation}
\theta_{\text{SHA}} = \frac{2e}{\hbar} \frac{\sigma_{xy}^{\text{SHC}}}{\sigma_{xx}}
\label{SHA}
\end{equation}
Where   $\sigma_{xy}^{SHC}$is the spin Hall conductivity (SHC) and  $\sigma_{xx}$ is the longitudinal charge conductivity.  In the present work, we consider the intrinsic SHE. The SHC was calculated via mapping DFT Hamiltonian to an effective Wannier Hamiltonian using Wannier90 approach and by using Kubo’s formula given by \cite{pizzi2020wannier90,qiao2018calculation},
\begin{equation}
\sigma_{xy}^{{SHC}} (E) = -\frac{e^2}{\hbar} \frac{1}{V N} \sum_{n k} \Omega_{n \mathbf{k}, x y}^{\text {spin,z}}(E)f_{nk}
\label{SHC}
\end{equation}
Here summation over $n$ is performed over all occupied bands, and N is the number of $k$-points in the first Brillouin zone. $f_{nk}$  is the Fermi-Dirac distribution function. The spin Berry curvature is defined by \cite{pizzi2020wannier90,qiao2018calculation},

\begin{equation}
\Omega_{n \mathbf{k}, x y}^{\text {spin,z}}(E)=\hbar^2 \sum_{m \neq n} \frac{-2 \operatorname{Im}\left[\left\langle\psi_{n \mathbf{k}}\right| \frac{2}{\hbar} \hat{j}_x\left|\psi_{m \mathbf{k}}\right\rangle\left\langle\psi_{m \mathbf{k}}\right| \hat{v}_y\left|\psi_{n \mathbf{k}}\right\rangle\right]}{\left(\epsilon_{n \mathbf{k}}-\epsilon_{m \mathbf{k}}\right)^2-(E+i \eta)^2}
\label{spin-berry}
\end{equation}
The longitudinal charge conductivity was calculated using the BOLTZWANN module \cite{pizzi2014boltzwann}. $\hat{j}_x^s$ is the spin-current operator given by $\hat{j}_x^s=\frac{1}{2}\{\hat{s_z},\hat{v_x}\}$ 
 
The calculated SHA ($\theta_{\text{SHA}}$) for the MoS$_2$/CoFeB is 0.022  while for MoS$_2$/Cu/CoFeB the corresponding value is 0.043. This supports the experimental claim that Cu effectively decouples MoS$_2$ from the magnetic exchange interaction with CoFeB, preserving the spin-orbit coupling (SOC) properties of MoS$_2$. Our theoretical results further validate the assertion that the intrinsic SHE, a key mechanism for spin-charge conversion in such systems, is augmented by interfacial modifications.

In Fig.\ref{fig:figure-4}(a) we show the atomic structure of the MoS\(_2\)/Cu/CoFeB heterostructure under consideration, while in the Fig.\ref{fig:figure-4} (b) we plot the spin Hall conductivity  ($\sigma_{xy}^{\text{SHC}}$) as a function of energy relative to the Fermi energy (\(E - E_F\)) for the two configurations. The red curve represents the MoS\(_2\)/CoFeB system, while the blue curve corresponds to the MoS\(_2\)/Cu/CoFeB system. Both curves exhibit oscillatory behavior, but the blue curve shows more periodic peaks than the red curve indicating enhanced spin Hall conductivity in the presence of the Cu spacer. In Fig.\ref{fig:figure-4} (c) we show the longitudinal charge conductivity ($\sigma_{xx}$) is plotted as a function of the chemical potential (\(\mu\)) for the two configurations.

From the denominator of Eq. \ref{spin-berry}, it is evident that peaks in the spin Hall conductivity (Fig. \ref{fig:figure-4}(b)) are associated with avoided crossings in the band structure near the Fermi level, which directly influence the spin Berry curvature due to SOC effects. In the MoS$_2$/CoFeB system, proximity-induced exchange interactions at the interface modify the electronic structure of MoS$_2$, partially suppressing its intrinsic SOC. The irregular peaks in the spin Hall conductivity suggest that the SOC effect is non-uniform and influenced by CoFeB's magnetic properties. Introducing a Cu interlayer between MoS$_2$ and CoFeB decouples MoS$_2$ from these magnetic interactions, resulting in more symmetric peaks in the spin Berry curvature for the MoS$_2$/Cu/CoFeB system. This results in enhanced spin Hall conductivity and reduced interfacial scattering in the Cu-intercalated system. This can be further understood by our calculation of magnetic anisotropy energy (MAE). Indeed, our calculated MAE values strongly support the discussion of spin Berry curvature and spin Hall conductivity behavior. The higher MAE for the  MoS$_2$/Cu/CoFeB (9.83$\times$10$^6$ erg/cm$^3$) system compared to  MoS$_2$/CoFeB (7.2$\times$10$^6$ erg/cm$^3$) aligns well with the enhanced preservation of SOC and reduced proximity-induced effects in the Cu-intercalated structure. This increased MAE reflects the effective decoupling achieved by the Cu interlayer, allowing MoS$_2$'s intrinsic SOC to dominate while mitigating proximity-induced effects from CoFeB. The consistency between the MAE calculations and the observed spin transport properties underscores the critical role of interface engineering in optimizing spintronic device performance.

\section{Conclusion}
We achieved a substantial enhancement in spin-to-charge conversion efficiency, with an increase of an order of magnitude, through Cu intercalation in the MoS$_2$/CoFeB system. Domain imaging confirmed that Cu effectively decouples MoS$_2$ from the proximity effects of CoFeB. The experimental findings are corroborated by theoretical calculations using the DFT-WANNIER90 framework, which reveal significant improvements in spin Berry curvature, spin Hall conductivity, and magnetic anisotropy energy. In summary, Cu intercalation preserves the intrinsic SOC of MoS$_2$ by mitigating magnetic exchange interactions with CoFeB, resulting in a marked enhancement of the spin Hall effect. This approach can further be explored to enhance the spin-to-charge conversion efficiency in other TMD materials like WSe$_2$, MoSe$_2$, MoTe$_2$ etc. These results highlight the potential of intercalation-based interface engineering strategies to enable the development of highly efficient spintronic devices.

\section{Acknowledgments}
AM, RC, SD, SS, KSR, and S. Bedanta thank the Department of Atomic Energy (DAE), Government of India, for the financial support via project with Sanct. No. 0803/2/2020/NISER/R\&D-II/8149 dated 16.07.2021). The authors acknowledge SERB project (CRG/2021/001245 dated 05.03.2022) for financial assistance. 
\nocite{kittel1948theory}
\nocite{brataas2002spin}
\nocite{mosendz2010detection}
\nocite{kresse1996efficiency}
\nocite{perdew1996generalized}
\nocite{daalderop1990first}
\nocite{liechtenstein1987local}
\bibliography{References} 
\end{document}